\documentclass[
	aps, prd, reprint,
	10pt, notitlepage, a4paper,
        floats, floatfix,
	amsmath, amssymb, amsfonts, 
	superscriptaddress,
	showpacs, showkeys,
	nofootinbib,
]{revtex4-1}

\def\paper{article}
\newcommand{\append}[1]{{#1}}

\usepackage{ifthen}
\newboolean{prd}
\setboolean{prd}{false}
\newboolean{arxiv}
\setboolean{arxiv}{true}

\newboolean{notprd}
\setboolean{notprd}{true}
\ifprd
\setboolean{notprd}{false}
\fi
\newboolean{notarxiv}
\setboolean{notarxiv}{true}
\ifarxiv
\setboolean{notarxiv}{false}
\fi

\usepackage{graphicx}
\usepackage[caption=false]{subfig}
\usepackage[usenames]{xcolor}

\ifnotprd
\xdefinecolor{mylinkcolor}{rgb}{0,0,0.5}
\usepackage[
	bookmarksnumbered, bookmarksopen, bookmarksopenlevel=2,
\ifnotarxiv
	breaklinks=true,
\fi
	colorlinks=true, filecolor=mylinkcolor, citecolor=mylinkcolor,
	linkcolor=mylinkcolor, urlcolor=mylinkcolor, menucolor=mylinkcolor,
]{hyperref}
\else
\newcommand{\texorpdfstring}[2]{{#1}}
\fi

\usepackage{wasysym}

\newcommand{\be}{\begin{equation}}
\newcommand{\ee}{\end{equation}}
\newcommand{\bea}{\begin{eqnarray}}
\newcommand{\eea}{\end{eqnarray}}

\def\vct#1{\mathbf{#1}}

\def\nnl{\nonumber \\ & \quad}

\def\nnlq{\nonumber \\ & \quad \qquad}

\DeclareMathOperator{\Order}{\mathcal{O}}

\allowdisplaybreaks

\begin{document}

\ifnotprd
\hypersetup{
	pdftitle={New perspectives on neutron star and black hole spectroscopy and dynamic tides},
	pdfauthor={Sayan Chakrabarti, Terence Delsate, Jan Steinhoff}
}
\fi

\title{New perspectives on neutron star and black hole spectroscopy and dynamic tides}

\author{Sayan Chakrabarti}
\affiliation{Centro Multidisciplinar de Astrof\'isica --- CENTRA, Departamento de F\'isica,
	Instituto Superior T\'ecnico --- IST, Universidade T\'ecnica de Lisboa, 
	Avenida Rovisco Pais 1, 1049-001 Lisboa, Portugal, EU}

\author{T\'erence Delsate}
\affiliation{Centro Multidisciplinar de Astrof\'isica --- CENTRA, Departamento de F\'isica,
	Instituto Superior T\'ecnico --- IST, Universidade T\'ecnica de Lisboa, 
	Avenida Rovisco Pais 1, 1049-001 Lisboa, Portugal, EU}
\affiliation{UMons, Universit\'e de Mons, Place du Parc 20, 7000 Mons, Belgium, EU}

\author{Jan Steinhoff}
\email[Corresponding author: ]{jan.steinhoff@ist.utl.pt}
\affiliation{Centro Multidisciplinar de Astrof\'isica --- CENTRA, Departamento de F\'isica,
	Instituto Superior T\'ecnico --- IST, Universidade T\'ecnica de Lisboa, 
	Avenida Rovisco Pais 1, 1049-001 Lisboa, Portugal, EU}
\affiliation{ZARM, University of Bremen, Am Fallturm, 28359 Bremen, Germany, EU}

\date{\today}

\begin{abstract}
We elaborate on a powerful tidal interaction formalism where
the multipole dynamics is kept generic and encoded in a linear response
function.
This response function is the gravitational counterpart of the atomic spectrum
and can become of similar importance with the rise of gravitational wave
astronomy.
We find that the internal dynamics of nonrotating neutron stars admit a
harmonic oscillator formulation yielding a simple interpretation of tides.
A preliminary investigation of the black holes case is given.
Our results fill the gap between Love numbers and dynamic tides. 
\end{abstract}


\maketitle

\section{Introduction}
Analytic models for gravitational interaction of compact objects in General
Relativity (GR) are plagued by potentially very complicated internal dynamics.
Recent progress on such tidal interactions is mostly focused on nondynamical
models \cite{Hinderer:2007, Damour:Nagar:2009:4, Binnington:Poisson:2009}, which
in particular can not account for oscillation modes.
This situation was already criticized and improved in
\cite{Maselli:Gualtieri:Pannarale:Ferrari:2012, Ferrari:Gualtieri:Maselli:2011}.
However, it may be difficult to extend this approach to objects other
than neutron stars (NS) and the internal dynamics is developed around
the Newtonian limit (but the adopted GR corrections seem to be sufficient
for most applications).
Here we devise a substantially more powerful tidal
interaction formalism based on an effective field theory (EFT) approach
\cite{Goldberger:Rothstein:2006}.
This approach was proposed in the context of black hole absorption
\cite{Goldberger:Rothstein:2006:2} and consists in
effectively replacing the extended object by a point particle comprising
dynamic covariant multipolar degrees of freedom (DOF). In this {\paper} the dynamics of the
multipoles is kept generic and encoded by a linear response function to external
tidal fields.

Motion of extended bodies in General Relativity (GR) has been subject to
question from the very beginning of the theory and gives the most important way
to test
gravity. Describing the dynamics of these objects is complicated
 and approximate methods have been developed, such as multipole
expansion schemes along the lines of Mathisson, Papapetrou, and Dixon
\cite{Mathisson:1937, Mathisson:2010, Papapetrou:1951, Dixon:1979}, between many
others. However, the definition of covariant compact-source multipoles in GR according
to Dixon is only useful for test bodies. The extension to self gravitating
objects is not fully understood, though it is clear that Dixon's
multipoles should be renormalized \cite{Harte:2012}. 

The adopted EFT approach implies a definition of covariant source multipoles of
self-gravitating objects in GR. This definition is implicit until an explicit
matching of the point-particle description to the actual extended object is
worked out. This is the main purpose of the present work, but also highly
nontrivial. (For instance, the determination of NS multipoles is already
quite subtle for nonperturbed stationary spacetimes \cite{Pappas:Apostolatos:2012},
see also \cite{Gurlebeck:2012}.)
This {\paper} substantially improves the situation for linear perturbations around
a static (nonlinear) background such as NS, Black Holes (BH),
White Dwarfs and likely even Boson Stars.

We illustrate our formalism with a simple Neutron Star (NS) model. The
tidal constants (Love numbers and yet undetermined constants) are easily
extracted from the response function.
It turns out that,  as long as linear perturbations are applicable, the
internal dynamics of NS admits a formulation in terms of harmonic oscillator
amplitudes \cite{Rathore:Broderick:Blandford:2002, Flanagan:Racine:2007,
Chakrabarti:Delsate:Steinhoff:2013:1}
similar to the Newtonian case, making the multipolar DOF
composite. This astonishing result leads to
simple and intuitive interpretations of tidal interaction in GR.
A full analysis of the black hole case is still in progress. Its
outcome is hard to foresee and thus for sure will bear surprises.


Although numerical simulations capture the nonlinear aspects of tidal
interactions, complementary analytic models stimulate invaluable
(at least qualitative) interpretations of the physical processes at hand.
An important aspect of our analytic dynamic tidal model is to naturally account
for resonances between external tidal fields and oscillation modes of the NS in
GR (see \cite{Shibata:1993, Reisenegger:Goldreich:1994, Lai:1994,
Kokkotas:Schafer:1995, Ho:Lai:1998, Flanagan:1998, Lai:Wu:2006, Flanagan:Racine:2007,
Press:Teukolsky:1977}). Such
resonances are of great importance. For instance, it was suggested recently that
the oscillations excited by these resonances can be strong enough to shatter the
NS crust, thus producing a weak short Gamma Ray Burst (GRB) \cite{Tsang:etal:2012}
(more precisely, a weak precursor to the main flare of the GRB produced by the
merger of the binary).
Besides such spectacular effects, resonances can of course leave more subtle,
but invaluable, imprints on the internal structure in the Gravitational Wave
(GW) signal. Numeric relativity simulations suggest that oscillations excited by
resonances can even be driven into the nonlinear regime and thus contribute
significantly to GW
\cite{Gold:Bernuzzi:Thierfelder:Brugmann:Pretorius:2011}.

The next revolution in GR will certainly arise from GW observatories like
Advanced LIGO and VIRGO. These detectors will begin its operation soon and
likely detect GW from binary NS mergers on a regular basis
\cite{Harry:LIGO:2010}.
Such GW signals encode a tremendous amount of information on the internal
structure of NS. This expectation is supported by recent numerical simulations,
which reveal imprints of the equations of state \cite{Bauswein:Janka:2011,
Sekiguchi:Kiuchi:Kyutoku:Shibata:2011:2} or the formation of a metastable
hypermassive NS \cite{Sekiguchi:Kiuchi:Kyutoku:Shibata:2011:1}.
Simultaneous detection of GW and GRB can provide for the first time
persuasive evidence for certain GRB scenarios 
\cite{Kiuchi:Sekiguchi:Shibata:Taniguchi:2010, Tsang:etal:2012}.

The present {\paper} is a continuation of our work in
\cite{Chakrabarti:Delsate:Steinhoff:2013:1} (on the Newtonian case) and we
adopt notations and conventions therein.

\section{Effective Action}
Our approach to account for the innumerable internal DOF of compact objects
follows along the lines of thermodynamics. The obstacle is to identify state
variables, which by definition describe the system on large scales (infrared,
IR). In the case of gravitational interaction, these state variables reduce
to the source multipole moments: At the same time they encode the IR
field and the motion \cite{Dixon:1979} of the object.

An effective point-particle action along the lines of
\cite{Goldberger:Rothstein:2006:2, Goldberger:Rothstein:2006, Goldberger:Ross:2009}
is most natural to implement covariant multipoles as macroscopic variables,
\begin{align}\label{Seff}
S_{\text{eff}} &= \int d \tau \left[ - m - \frac{1}{2} E_{ab} Q^{ab} + \dots \right] ,
\end{align}
where $m$ is the mass of the NS and $E_{ab}$ is the electric part of the Weyl tensor.
For simplicity, we only discuss the covariant electric type quadrupole $Q^{ab}$ here, but
 inclusion of other multipoles is straightforward, see
\cite[Eq.\ (1)]{Goldberger:Ross:2009}. The indices $a, b$ indicate the
spatial components in a local
Lorentz frame comoving with the NS. The worldline parameter $\tau$ is
the proper time here.

We consider linear perturbations of compact objects, so we expect a linear
response
of the quadrupole to the (quadrupolar) tidal field $E_{ab}$,
\begin{equation}\label{Fdef}
\tilde{Q}^{ab}(\omega) = - \frac{1}{2} \tilde{F}(\omega) \tilde{E}^{ab}(\omega) ,
\end{equation}
where the tilde denotes Fourier transformation from $\tau$ to $\omega$, and
$\tilde{F}$ is the linear response function (or propagator).
The main objective of the present
{\paper} is to determine $\tilde{F}$ from a matching procedure. 
As explained in \cite{Chakrabarti:Delsate:Steinhoff:2013:1}, from a
Taylor-expansion
\begin{equation}\label{Lovedef}
\tilde{F}(\omega) = 2 \mu_2 +i \lambda \omega + 2 \mu'_2 \omega^2 +
\Order(\omega^3) ,
\end{equation}
the tidal constants $\mu_2$ and $\mu'_2$ emerge.
The first parameter $\mu_2$ is related to the dimensionless (relativistic,
quadrupolar, 2nd-kind) tidal Love number $k_2 = 3 G \mu_2 / 2 R^5$, where
$R$ is the radius and $G$ is the Newton constant, in agreement with definitions in
\cite{Damour:Nagar:2009:4}.
Furthermore, $\mu'_2$ parametrizes the tidal response beyond the
adiabatic case. Though it was formally introduced in \cite{Bini:Damour:Faye:2012},
it was not determined numerically yet. It obviously comes out as a
byproduct within our approach.
The constant $\lambda$ is related to absorption
\cite{Goldberger:Rothstein:2006:2}, see also \cite{Porto:2007,
Kol:2008, Goldberger:Ross:Rothstein:2012} and for a non-EFT treatment see, e.g.,
\cite{Poisson:2005}. The time dependence of the mass parameter in the effective
action due to absorption is discussed in \cite{Goldberger:Ross:Rothstein:2012}.
The response function is analogous to the refractive index in optics, where
imaginary parts also encode absorption. This analogy enlightens
the matching procedure. Indeed, the phase shift between ingoing and
outgoing waves encodes the real part of the response (Love number/refractive
index) while the change in amplitude is due to absorption. But the nonlinear nature
of GR makes the interpretation of phase shifts more subtle.

We should stress that besides encoding all quadrupolar tidal constants in a single function
$F$, our approach can naturally accommodate the presence of oscillation modes
that are obviously missed by a Taylor expansion \eqref{Lovedef}. This
possibility was not discussed in \cite{Goldberger:Rothstein:2006:2}, where
the focus is on absorption.

Generic extensions of the point-mass action were considered in \cite{Bailey:Israel:1975}
and the resulting EOM were related to Dixon's results. This can readily
be applied to (\ref{Seff}).
Explicit expressions for the stress tensor in terms of Dirac delta
distributions can be found in \cite{Steinhoff:Puetzfeld:2009, Steinhoff:2011}.
However, it should be noted that the relation between Dixon's covariant multipole moments
and the covariant moments used in the action (\ref{Seff}) is more of a formal nature when
self-gravitating objects are considered.


It is straightforward to derive the Newtonian interaction potential for binaries
belonging to the effective action. Even the first post-Newtonian (PN)
correction for a generic quadrupole was already worked out
\cite{Vines:Flanagan:2010} (though not from the effective action; see
also \cite{Vines:Hinderer:Flanagan:2011} for the impact on GW).
However, the dynamics of the quadrupole was essentially left open and only
made explicit for the adiabatic case.
The present work fills this gap by providing a dynamical quadrupole model.
It should be stressed that even if the effective action is applied to Newtonian
or PN approximations, the response function $\tilde{F}$ encodes strong field
aspects of GR. This is the eminent advantage of the EFT approach.
PN interaction potentials including tidal coefficients were derived in
\cite{Damour:Nagar:2009, Vines:Flanagan:2010, Bini:Damour:Faye:2012}.




\section{Perturbed compact objects}
Without going into detail, we just mention here that perturbations of static
compact objects can be determined from a system of coupled ordinary differential
equations with the radial coordinate as the variable and the frequency $\omega$
entering as a parameter.
(The specific case of  nonrotating spherical
symmetric NS perturbation goes back
to \cite{Thorne:Campolattaro:1967,Thorne:1969}, for reviews see
\cite{Kokkotas:Schmidt:1999, Nollert:1999, Rezzolla:2003}.) 
In the exterior, the perturbation equations
are given by the famous Zerilli \cite{Zerilli:1970} or Regge-Wheeler (RW)
\cite{Regge:Wheeler:1957}
equations for electric- or magnetic-parity type perturbations, respectively.
These are the same equations that describe perturbations of Schwarzschild BH.
The Zerilli equation can be cast into the (simpler) RW form
\cite{Chandrasekhar:1975}, such that the discussion can be restricted to the
latter. Moreover, the RW equation possesses analytic series solutions
\cite{Mano:Suzuki:Takasugi:1996}, see also \cite{Leaver:1986:1,
Sasaki:Tagoshi:2003}, which are central for the present
work.
Our approach consists in solving the perturbation equations numerically in
the interior and connecting to the analytic vacuum solutions by
imposing appropriate boundary conditions at the surface.

As the RW equation is a second order homogeneous differential equation, its
generic solutions can be represented by a linear combination of two independent
solutions.
In \cite{Mano:Suzuki:Takasugi:1996},
the pair of analytic solutions $X_0^{\nu}$ and
$X_0^{-\nu-1}$ represented by series of Gauss Hypergeometric Functions converges
at the BH horizon but not at spatial infinity, while
for $X_C^{\nu}$ and $X_C^{-\nu-1}$ (series of Confluent Hypergeometric
Functions) it is the other way around. Here $\nu$ is
the renormalized angular momentum, which is fixed by requiring convergence of the
analytic series solutions \cite{Mano:Suzuki:Takasugi:1996}.
We review the needed elements of \cite{Mano:Suzuki:Takasugi:1996} in Appendix \ref{MST}.
The solutions can be matched as
\begin{equation}
X_0^{\nu} = K_{\nu} X_C^{\nu} , \quad
X_0^{-\nu-1} = K_{-\nu-1} X_C^{-\nu-1} ,
\end{equation}
where $K_{\nu}$ is given by \cite[Eq.\ (4.2)]{Mano:Suzuki:Takasugi:1996}
or (\ref{KMST}).

In the absence of dissipation, $\tilde{F}$ should be real so it
is natural
to work with manifestly real quantities. Before proceeding, we thus
introduce normalization constants $N_{\nu}$ such that the analytic solutions
$X_N^{\nu} := N_{\nu} X_C^{\nu}$ are real. We also require that the
asymptotic amplitude is 1, i.e., $N_{\nu}$ is uniquely defined by the
requirement
\begin{equation}
X_N^{\nu} \stackrel{r_* \rightarrow \infty}{\sim}
        1 \times \cos (\omega r_* + \text{const}) \in \mathbb{R} ,
\end{equation}
where $r_* = r + 2 M \log (r / 2 M - 1)$ is the tortoise coordinate,
$r$ is the Schwarzschild radial coordinate, and we identify $M = G m$.
It is straightforward to work out an explicit series representation for $N_{\nu}$
from the formulas provided in \cite{Mano:Suzuki:Takasugi:1996,
Leaver:1986:1}. The result is shown in Appendix \ref{normMST}.
Now the RW function $X$ in the exterior can be decomposed as
\begin{equation}\label{Xnum}
X = A_1 X_N^{\nu} + \epsilon^4 A_2 X_N^{-\nu-1} .
\end{equation}
The main numeric result needed for our investigation is encoded in the
amplitudes $A_1$ and $A_2$. The introduction of $\epsilon^4$ is suggested by
an analysis of $K_{\nu}$ and $K_{-\nu-1}$ for small $\epsilon$.

We checked our implementation of the analytic solutions
\cite{Mano:Suzuki:Takasugi:1996} against a direct integration method
\cite{Chandrasekhar:Detweiler:1975}.



\section{EFT Calculation}
The generic idea is to replace the compact object by an effective source encoded
by
the action (\ref{Seff}), such that the RW function $X$ at large distance
(in the IR) is reproduced. This singular source can be expressed
in terms of Dirac delta distributions. In the effective theory we therefore
need to solve an inhomogeneous RW equation
\begin{equation}\label{inhRW}
\frac{d^2 X}{d r_*^2} + \left[ \left(1-\frac{2 M}{r} \right)
\frac{l (l+1) - \frac{6 M}{r}}{r^2} + \omega^2 \right] X = S[X] ,
\end{equation}
where $l$ is the angular momentum quantum number.
The source term $S$ can be derived by projecting the stress tensor following
from the action (\ref{Seff}) onto tensor spherical harmonics, completely analogous
to a point-mass source \cite{Zerilli:1970:2, Davis:Ruffini:Press:Price:1971}.
Explicit expressions for the quadrupole case $l=2$ are supplied in Appendix \ref{effS}.
The principle is the same for other values of $l$.

As the distributional source $S$ should mimic the compact object, it must be located
at $r=0$. However, due to the (regular) singularity at $r=2M$, the
inhomogeneous RW equation (\ref{inhRW}) then does not seem to make
sense. This problem is resolved by understanding the solutions as
expanded in the post-Minkowskian expansion parameter $M$ or $\epsilon = 2 M \omega$.
Expanding the solutions in $\epsilon$ is subtle due to various poles arising from
Gamma Functions. If one keeps $l$ generic and performs the limit $l \rightarrow 2$ after
the expansion, one ends up with a different set of solutions to the RW equation denoted
$X_{\delta_l}^{\nu}$ and $X_{\delta_l}^{-\nu-1}$, where $\delta_l = l-2$ represents the
deviation from the quadrupole case $l=2$. It holds
\begin{align}
& X_N^{\nu} = X_{\delta_l}^{\nu} \left[ 1 + \frac{7 \epsilon ^6}{1605 \delta_l} \right]
+ X_{\delta_l}^{-\nu-1} \bigg[ -\frac{7 \epsilon ^5}{1605}
   - \bigg( \frac{7}{3210 \delta_l^2} \nnl
   +\frac{1}{450 \delta_l}+\frac{10548481}{1442574000}\bigg) \epsilon ^7 \bigg]
+ \Order(\epsilon^8, \delta_l) , \label{Xregular} \\
& X_N^{-\nu-1} = - X_{\delta_l}^{-\nu-1} \bigg[\frac{107}{210} \epsilon
    + \bigg(\frac{107}{420 \delta_l^2}+\frac{2165423}{18522000}\bigg)
   \epsilon^3\bigg] \nnl
   + X_{\delta_l}^{\nu} \bigg[1+\bigg(\frac{107}{210 \delta_l}-\frac{11449}{88200}\bigg) \epsilon^2\bigg]
   + \Order(\epsilon^4, \delta_l) .
\end{align}
Notice that the $\delta_l$-poles in the coefficients are canceled by poles contained in the
solutions $X_{\delta_l}^{\nu}$ and $X_{\delta_l}^{-\nu-1}$. Despite these complication
we work with $X_{\delta_l}^{\nu}$ and $X_{\delta_l}^{-\nu-1}$. The reason is that by keeping
$l$ generic one can easily identify the terms produced by the source $S$.

Still the solutions may be singular at $r=0$, so regularization techniques are needed to handle the delta
distributions contained in the source $S$. We choose a smooth ultraviolet "cutoff" in the
form of a Riesz-kernel representation, see, e.g., \cite{Damour:Jaranowski:Schafer:2008:2}.
Furthermore, as the calculation of the source $S$ for generic $l$ is not an easy one, we use a
more ad hoc approach. We take $S$ for $l=2$ only and multiply the Riesz kernel by $r^{-\delta_l}$
to augment it with a fractional multipole character. Finally we represent $\delta(\vct{r})$ by
\begin{equation}\label{Riesz}
\delta(\vct{r}) = ( r c_l )^{-\delta_l} \, "\!\lim_{\delta \rightarrow 0}\!" \frac{\Gamma(\frac{d-\delta}{2})}{\pi^{d/2} 2^{\delta} \Gamma(\frac{\delta}{2})}
	\mu_0^{\delta} r^{\delta-d} ,
\end{equation}
where $d=3$ is the number of spatial dimensions. $\mu_0$ and $c_l$ are arbitrary parameters
of unit inverse length formally introduced to make the expression dimensionally
correct. The limit $\delta \rightarrow 0$ is understood to be taken in final expressions.

We are going to construct the solution to the inhomogeneous equation from the solution
to the homogeneous one using the standard method of variation of
parameters/constants. This allows us to reinterpret the analytic solutions $X_{\delta_l}^{\nu}$
and $X_{\delta_l}^{-\nu-1}$, which converge in the IR, as the ones belonging
to the EFT. So the ansatz reads
\begin{equation}\label{XEFT}
X = C_1 X_{\delta_l}^{\nu} + C_2 X_{\delta_l}^{-\nu-1} ,
\end{equation}
where $C_1$ and $C_2$ are yet unknown functions of $r$.
If $C_1$ and $C_2$ were constants, then this would just be a solution
of the homogeneous equation.
It is an elementary result that
\begin{subequations}\label{varC}
\begin{align}
C_1(r') &= - \int_0^{r'} \frac{S X_{\delta_l}^{-\nu-1}}{W_*} \frac{d r_*}{d r} d r + H_1 , \\
C_2(r') &= \int_0^{r'} \frac{S X_{\delta_l}^{\nu}}{W_*} \frac{d r_*}{d r} d r + H_2 ,
\end{align}
\end{subequations}
where $H_1$ and $H_2$ are yet undetermined integration constants and $W_*$
is the Wronskian w.r.t.\ the tortoise coordinate,
\begin{equation}\label{Wrons}
W_* = X_{\delta_l}^{\nu} \frac{d X_{\delta_l}^{-\nu-1}}{d r_*} - X_{\delta_l}^{-\nu-1} \frac{d X_{\delta_l}^{\nu}}{d r_*} .
\end{equation}
By virtue of Abel's identity, this Wronskian is actually a constant and can be
evaluated at $r=\infty$ by analyzing the asymptotic behavior of the analytic solutions,
see (\ref{WronsMST}).
This procedure to solve the RW equation with a delta source has some similarities to
the construction of the gravitational Green function, which was also obtained from
analytic solutions very recently \cite{Casals:Ottewill:2012:2}. It can be interesting to
further study this connection in the future.

Next one must constrain the integration constants $H_1$ and $H_2$. The
solution they represent is not allowed to correspond to further delta-type sources
at $r=0$. This argument leads to $H_2=0$, as the sourced terms are proportional
to $r^{-l-1} \sim X_{\delta_l}^{-\nu-1}$, while $X_{\delta_l}^{\nu} \sim r^l$.
If we would set $l=2$ in the very beginning, then one must identify the constraint
on $H_1$ and $H_2$ by iteratively solving the RW equation in $\epsilon$. However,
the coefficients in the numeric solution (\ref{Xnum}) scale
differently by 4 orders in $\epsilon$, so one must iterate to the same order
in $\epsilon$ just for the leading order result. This is
the reason for keeping $l$ generic here, at the cost of introducing poles in
$\delta_l$ into the calculation. Dimensional regularization would solve this
problem in a similar manner, see \cite{Kol:Smolkin:2011}. Interestingly in the
static limit the dependence on the dimension can be absorbed into $l$ \cite{Kol:Smolkin:2011}.

It should be emphasized that the lower integration bound in (\ref{varC}) is $r=0$,
which is only possible if the integrand is understood as expanded in $\epsilon$.
Then the singularity of $d r_* / d r$ and the oscillatory behavior of the
analytic solutions at the horizon are removed. Here nonconvergence of $X_{\delta_l}^{\nu}$ and
$X_{\delta_l}^{-\nu-1}$ at the horizon is actually not a flaw, but a feature.
Furthermore, Eqs.\ (\ref{varC}) only start to depend on the upper integration bound $r'$ at
linear order in $\delta$, so $C_1$ and $C_2$ are actually constant for $\omega \rightarrow 0$.

Finally, the quadrupole components entering the source $S$ must be determined
according to (\ref{Fdef}). This requires to evaluate the tidal field $\tilde{E}^{ab}$
at $r=0$, or
\begin{equation}\label{Qren}
\tilde{Q}^{ab}(\omega) = - \frac{1}{2} \tilde{F}(\omega)
        \int \tilde{E}^{ab}(\vct{r}, \omega) \delta(\vct{r}) \, d^3 \vct{r} ,
\end{equation}
where again the Riesz kernel provides the necessary regularization and
the integrand is expanded in $\epsilon$ before evaluation.
Note that $\tilde{E}^{ab}(\vct{r}, \omega)$ follows from the RW master
function (\ref{XEFT}) and the background.
See Appendix \ref{quadcomp} for its representation in RW gauge.

\section{Results}
Comparing the numerically obtained solution (\ref{Xnum}) with the effective
one (\ref{XEFT}) leads to two conditions.
These are solved for the yet undetermined quantities $H_1$
and $\tilde{F}$ in terms of $A_1$ and $A_2$, completing the computation.
At the end of the day, one arrives at
\begin{align}\label{Fresult}
& \frac{3 G}{4 M^5} \tilde{F} =
   -\frac{428 A_2}{7 A_1} \bigg\{ 1 - \epsilon ^2 \bigg[
   \frac{33054269}{9437400} \nnl
   + \frac{107}{105} \bigg( \frac{1}{\delta} - \log \frac{2 \omega }{\mu_0} \bigg)
   \bigg] \bigg\}
-\frac{56}{107} \bigg\{ 1 + \epsilon ^2 \bigg[
   \frac{13138723}{18874800} \nnl
   - \frac{107}{105} \bigg[ \frac{1}{\delta} - \frac{1}{2 \delta_l} + \gamma_E
        - \log \frac{\mu_0}{c_l} \bigg]
   \bigg] \bigg\}
   + \Order(\epsilon^4) .
\end{align}
Let us insist again that though this seems to be an expansion in $\epsilon$, 
the numerical quantity $A_2 / A_1$ can still capture strong field effects from the interior.
We define a renormalized $\tilde{F}_{\text{MS}}$ by dropping the poles in
$\delta$ and $\delta_l$, analogous to
minimal subtraction (MS) in dimensional regularization.

We apply our method to the astrophysically most relevant case of NS.
The used system of differential equations is derived in Appendix \ref{NSpeturb}.
We use a simple polytropic equation of state (EOS) with index $1$ for the
nuclear matter.
The results presented below are for a NS with $m = 1.27 m_{\astrosun}$ and
$R = 8.85 \, \text{km}$. The complex quasi-normal mode frequencies for this
specific
NS model were reported in \cite{Kokkotas:Schmidt:1999} and our numeric
implementation
reproduces them very well (except for the damping of the curvature modes).

\begin{figure}
\includegraphics{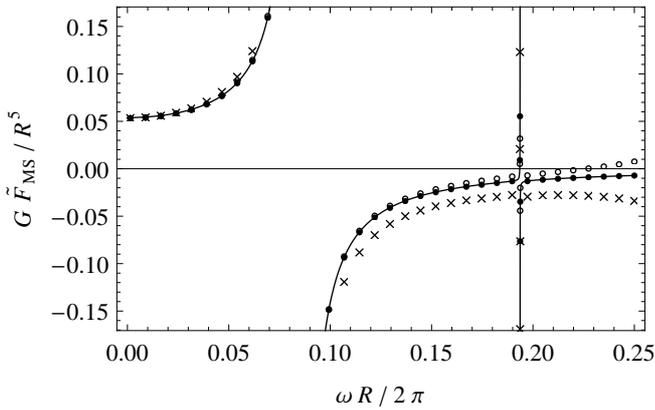}
\caption{Propagator of the quadrupole in minimal subtraction (MS). The dots are some
selected numeric data points, the solid line is the fit (\ref{fitF}) and (\ref{fitmu}).
For the circles the logarithmic scale dependence was ignored ($\mu_0 \sim \omega$)
and for the crosses only the leading order in (\ref{Fresult}) was taken
into account. This shows that the $\epsilon^2$ corrections in (\ref{Fresult}) are
essential for a good fit, while the corrections from $\mu_0$ start to contribute
only beyond the f-mode.\label{propGR}}
\end{figure}

An excellent fit for
$\tilde{F}_{\text{MS}}$ for this NS model turns out to be
\begin{equation}\label{fitF}
\frac{G \tilde{F}_{\text{MS}}}{R^5} \approx \frac{q_f^2}{R^2 (\omega_f^2 - \omega^2)}
   + \frac{q_p^2}{R^2 (\omega_p^2 - \omega^2)} ,
\end{equation}
provided we also fit the renormalization scale $\mu_0$.
We numerically generated a set of 350 data points (with higher density near the
poles)
and the fit deviates from all of them by at most 2\% (see Fig.\
\ref{propGR}).
The optimal fit parameters are given by
\begin{gather}
\omega_f R = 2 \pi \, 0.0851 , \quad
q_f = 1.98 \times 10^{-2} , \label{fitmu} \\
\omega_p R = 2 \pi \, 0.194 , \quad
q_p = 9.1 \times 10^{-4} , \quad
\mu_0 R = 0.6 . \nonumber
\end{gather}
It is straightforward to infer the tidal constants defined
by (\ref{Lovedef}).
It should be noted that all fit parameters are essentially
independent of $c_l$, which we varied from $\omega$ to $\mu_0$.

It is remarkable that the quadrupole propagator can be approximated by a sum of
response functions of harmonic oscillators, just like in the Newtonian case
\cite{Chakrabarti:Delsate:Steinhoff:2013:1}.
The relativistic case thus seems to admit an amplitude
formulation \cite{Rathore:Broderick:Blandford:2002, Flanagan:Racine:2007}
analogous to the Newtonian case.
The a priori very complex internal dynamics of the NS
is then approximated just by a set of harmonic oscillators, which are the more
fundamental effective DOF composing the dynamical quadrupole.
The constants $q_f R^3/G$ and $q_p R^3/G$ can be understood as GR versions
of the overlap integrals. Resonances are quantitatively described by
forced harmonic oscillators. An extension of this mechanical picture to
nonlinear oscillators and/or mode coupling can offer a demonstrative
phenomenological way to model even more realistic situations.

The frequency dependent Love number introduced recently in
\cite{Maselli:Cardoso:Ferrari:Gualtieri:Pani:2013},
where the formalism in \cite{Maselli:Gualtieri:Pannarale:Ferrari:2012,
Ferrari:Gualtieri:Maselli:2011} was used, should be related to
our response function in the low frequency regime and a comparison
is most interesting due to the very different setup (single object in
perturbation theory vs.\ complete binary). It is further an interesting question
whether the expansion (\ref{Lovedef}) including the tidal coefficient
$\mu'_2$ introduced in \cite{Bini:Damour:Faye:2012} is enough to find
agreement within the regime where a comparison is possible.

It is well known that the f-mode properties (frequency) basically just depend
on the mean density of the NS \cite{Kokkotas:Schmidt:1999}. To a good
approximation this should also be true for the tidal response of the quadrupole,
as we find here that it is largely dominated by the f-mode (at least for the adopted NS model).
This aspect may be related to the universal relations discussed
in \cite{Yagi:Yunes:2013:1, Yagi:Yunes:2013:2, Maselli:Cardoso:Ferrari:Gualtieri:Pani:2013}.

Also the BH case can be readily investigated using (\ref{Fresult}). One can even
work out analytic formulas for $A_1$ and $A_2$ from
$X^{\text{BH}} \propto X_0^{\nu} + X_0^{-\nu-1}$ \cite{Mano:Suzuki:Takasugi:1996},
reading
\begin{equation}
A_1 = \frac{K_{\nu}}{N_{\nu}} , \quad
\epsilon^4 A_2 = \frac{K_{-\nu-1}}{N_{-\nu-1}} .
\end{equation}
The nontrivial character of this analytic result becomes apparent once
(\ref{KMST}), (\ref{NMST}), (\ref{N1MST}), and (\ref{N2MST}) are inserted.
Because of the absorption due to the horizon one can not
expect poles for $\omega \in \mathbb{R}$ like in the NS case. This makes
the analysis more complicated, as the whole complex plane must be considered.
However, one can immediately obtain an expansion of $\tilde{F}$ in $\epsilon$,
\begin{equation}\label{FBH}
\begin{split}
\frac{G \tilde{F}^{\text{BH}}_{\text{MS}}}{(2 M)^5} &= \frac{i \epsilon }{45}
   + \epsilon^2 \bigg[ \frac{3486611}{54096525}
   - \frac{1}{45} \log (2 M c_l) \bigg]
   + \Order(\epsilon^3) .
\end{split}
\end{equation}
This implies that the BH Love number $\mu_2$ vanishes, in agreement with
the findings in \cite{Fang:Lovelace:2005, Damour:Nagar:2009:4,
Binnington:Poisson:2009, Kol:Smolkin:2011}. Unfortunately the
unspecified parameter $c_l$ can substantially influence the next
order tidal coefficient $\mu_2'$.
This makes clear that for a rigorous investigation one should first
redo the EFT calculation within a better regularization method
like dimensional regularization.

But it should be stressed that the Love number is independent of
$c_l$. In fact, it is possible to obtain the leading order of
(\ref{Fresult}) by setting $\delta_l=0$ throughout the computation
using a shortcut.
Though $l$ is not available to identify the constraint on $H_1$ and $H_2$,
a simple argument can be given at leading order. $H_1$ and $H_2$ must correspond to
the linear combination of $X_N^{\nu}$ and $X_N^{-\nu-1}$
that cancels the $z^{-2}$-term in the solution at orders $\epsilon^4$ and
$\epsilon^5$. Unfortunately at order $\epsilon^6$ [corresponding to $\epsilon^2$
in (\ref{Fresult})] this approach breaks down due to $z^{-2} \log z$ contributions
to the solution, which can only be interpreted by an iteration of the field
equations. Still it is highly desirable to reach the next to leading order,
as illustrated by Fig.\ \ref{propGR}.

\section{Conclusions}
The validity of our results is supported in many ways. First of all,
by comparing the adiabatic limit of the analytic solutions used here against the analytic
zero-frequency solutions used in \cite{Damour:Nagar:2009:4} one can show that
the definition of $\mu_2$ in \cite{Damour:Nagar:2009:4} agrees with the definition
through (\ref{Lovedef}) and (\ref{Fresult}).
Second, the term linear in $\epsilon$ in (\ref{FBH}) was already derived in
\cite{Goldberger:Rothstein:2006:2} and agrees with our findings. Finally,
an intermediate result is the dependence of quadrupole components $\tilde{Q}^{ab}$
from (\ref{Qren}) on the scale $\mu_0$, which in fact agrees
with the beta function found in \cite{Goldberger:Ross:2009} using dimensional 
regularization (if it is assumed that $c_l$ is independent of $\mu_0$).

But unfortunately our current results depend on another parameter
$c_l$ with unclear interpretation. This situation can be improved
by applying dimensional regularization to the EFT calculation, which
is also most useful for applications to post-Newtonian theory.
But we expect our results to be good approximations for the dimensional
regularized ones (the matching scale $\mu_0$ will be slightly different).
Still this calls for a clearer connection between our formalism and
Ref.\ \cite{Goldberger:Ross:2009} (where also the background Schwarzschild
metric is generated within the EFT).
At the same time, higher multipoles (including magnetic/axial) should be treated as well.

However, these current problems with the regularization method play no
role for the static limit. Thus no ambiguities for
the definition of the Love numbers emerge. The predictions for the
RW function from the EFT are "simply" matched to the numeric results,
as in \cite{Kol:Smolkin:2011} for the black hole case in dimensional
regularization.
Consequently there is no need to interpret the definition of $k_2$ as relative to
the BH case in the current approach.
Furthermore, our computation is based on the exterior solution, so it
is applicable to arbitrary (nonrotating) compact objects.



Another obvious next step is an application of our method to more
realistic NS models. Besides realistic EOS, an investigation
of the NS crust is most promising due to a possible connection
to precursors flares in short GRB \cite{Tsang:etal:2012} by a shattering of the
crust.

Further, realistic NS are rotating. Neutron star modes can
become unstable in the rotating case (including the f-mode
\cite{Gaertig:Glampedakis:Kokkotas:Zink:2011}), which can
also give rise to violent astrophysical processes.
An extension of the present method can be tried within a slow
rotation approximation, see, e.g.,
\cite{Pani:Cardoso:Gualtieri:Berti:Ishibashi:2012,
Pani:Cardoso:Gualtieri:Berti:Ishibashi:2012:2} and references therein.
The r-modes of rotating NS are of particular interest for
resonances \cite{Flanagan:Racine:2007}.
 

Analytic predictions for GW including tidal effects
from an Effective One-Body (EOB) approach agree even quantitatively with numeric simulations
\cite{Baiotti:etal:2010, Damour:Nagar:2009}. Yet the tides are modeled by Love numbers
and absorption \cite{Nagar:Sarp:2011} only. Inclusion of the dynamical multipole response
function into the EOB formalism is expected to establish the impact of
resonances between NS modes and orbital motion on GW in a
reliable manner.

Our approach can probably also be evolved into a method for
finding oscillation modes. In the conservative case, modes can be
found by "just" integrating the perturbation equations for real frequencies.
Estimates for the damping times can be obtained using the quadrupole formula
(and basically correspond to the overlap integrals).
However, if the modes are damped by, e.g., dissipative effects in the nuclear
matter or mode coupling, then the poles of the response function
should have a nonvanishing imaginary part. This also illustrates that our
approach separates properties of the star from effects due to the surrounding
spacetime, which is nontrivial as the background is nonlinear and does not admit
superposition arguments.

The BH case is largely left unexplored for now. At the same time,
the prospects are fascinating. If a mechanical oscillator model for
BH can be formulated, then one can further elaborate on the thermodynamic
analogy of these macroscopic DOF.
This can lead to insights on macroscopic concepts like BH entropy and
temperature from an EFT point of view.
Analogies to the AdS/CFT correspondence discussed in \cite{Goldberger:Rothstein:2006:2,
Goldberger:Rothstein:2006:3} can probably be made more explicit, too.
The highly damped modes and branch cuts (eventually introduced by
nonanalytic terms like $\log \epsilon$) can be difficult to handle.
Extension to the case of rotating BH should be almost straightforward, as
the perturbation master equations \cite{Teukolsky:1972}, the
appropriate analytic solution \cite{Mano:Suzuki:Takasugi:1996:2},
and the effective action \cite{Porto:2007} are readily available.
The Love number of rotating BH should even come out unambiguously
if the current (improvable) regularization method is applied.

For an application to BH scattering a simplistic fit of the response
function can be accurate enough. This can lead to interesting connections
to, e.g., the scattering thresholds discussed in
\cite{Berti:etal:2010, Sperhake:etal:2009}.
Finally, if the present method is applied to perturbations of massive
scalar fields around rotating black holes, the floating orbits existing
for extreme mass ratios \cite{Cardoso:Chakrabarti:Pani:Berti:Gualtieri:2011}
can possibly be constructed for comparable mass binaries using PN methods.


\acknowledgments
We acknowledge fruitful discussions with P. Pani and V. Cardoso.
We are further grateful for useful comments from M.\ Casals,
L.\ Gualtieri, N.\ G{\"u}rlebeck, B.\ Kol, and A.\ Maselli.
This work was supported by DFG (Germany) through project STE 2017/1-1 and STE 2017/2-1,
FCT (Portugal) through projects PTDC/CTEAST/098034/2008 and PTDC/FIS/098032/2008 and CERN through project 
CERN/FP/123593/2011.

\ifnotprd
\bibliographystyle{utphys}
\fi

\ifarxiv
\providecommand{\href}[2]{#2}\begingroup\raggedright\endgroup

\else
\bibliography{QNM_resonance_PN}
\fi

\append{
\begin{widetext}
\appendix
\newpage

\section{Formulas and Implementations}

\subsection{NS Perturbation Equations\label{NSpeturb}}

\subsubsection{Preliminaries}
A static spherically symmetric star configuration is described by a metric
\be
ds^2_0 = -f(r) dt^2 + b(r) dr^2 + r^2 d\Omega^2,
\ee
where $d\Omega$ is the line element on the unit sphere, and a perfect fluid
with stress tensor is given by
\be
T_{\mu\nu} = (\rho + P) u_{\mu} u_{\nu} + P g_{\mu\nu},
\ee
where $\rho$ is the density, $P$ is the pressure, $g_{\mu\nu}$ is the metric
components, $u^{\mu}$ is the four-velocity such that $u_\mu u^\mu = -1$,
and Greek indices belong to the spacetime coordinate basis.
In the NS case, the description is usually completed by a barotropic EOS
relating $P$ and $\rho$ (neglecting temperature), according to
\be
\rho(r) = \bar\rho(P(r)),
\ee
for a given function $\bar\rho$. In this {\paper} we considered a polytropic
EOS defined by 
\be
P = K \rho^{\frac{n+1}{n}}.
\ee

The dynamics of the compact object is then given by the Einstein equations and
the conservation equation of the stress tensor
\be
R_{\mu\nu} - \frac{1}{2}R g_{\mu\nu} = 16\pi G T_{\mu\nu},\
\nabla_\mu T^{\mu\nu} = 0.
\ee

The even parity metric perturbations around a spherically symmetric background are given in
the Regge-Wheeler gauge by
\be
ds^2_1 = -  f(r) h_0(x^\mu) dt^2 + 2 i \omega h_1(x^\mu)dr dt + 
 b(r) h_2(x^\mu)dr^2 +  r^2  k(x^\mu)d\Omega^2,
\ee
where the functions $h\in\{h_0,h_1,h_2,k\}$ depend on the coordinates $x^\mu$
according to $h = h(r) \exp(i\omega t) Y_{lm}(\Omega)$, $Y_{lm}$ being the
scalar spherical harmonics.
The total metric then reads $ds^2 = ds^2_0 + ds^2_1$,
where the subscripts 0 and 1 denote the background and perturbation, respectively.
The perturbations to the matter fields are given by
\be
P =P_0(r) + P_1e^{-i\omega t} Y_{lm}(\Omega) ,\ \rho =
\rho_0(r) + \rho_1(r)e^{-i\omega t} Y_{lm}(\Omega),\ u^\mu =
u_0^\mu(r) + u_1^\mu(r,\theta,\varphi)e^{-i\omega t}.
\ee

Given an equation of state, the perturbation to the density is given in terms of
the perturbation to the pressure:
\be
\rho_1 = \frac{d \bar\rho}{dP_0}P_1 = \frac{P_1}{c_s^2},
\ee
which defines the speed of sound $c_s$.

In the following, it will be useful to introduce the function $U$ such that
\be
P_1 =\left( U -\frac{h_0}{2}\right)(\rho_0 + P_0). 
\ee

The solution for $u_1^\mu$ is given in terms of the function $U$ and of the
metric perturbation
\be
-\sqrt{f(r)} u_1^\mu dx_\mu = 
\left(\frac{h_0(r)}{2} dt + \frac{\omega h_1(r)-i f(r) U'(r)}{\omega
b(r)}dr\right)Y_{lm}(\Omega) + \frac{i f(r) U(r) }{r^2
\omega }\nabla^i Y_{lm} d\Omega_ i ,
\ee
where $d\Omega_i = (d\theta,d\varphi)$.

\subsubsection{Master Equations}
The perturbed equations can be solved for $h_0,h_2$ and their derivatives,
leaving three ordinary differential equations in the interior of the compact
object
\bea
&&U''=- \bigl\{b^3 f k \bigl(r + 8 \pi r^3 \
P\bigr)^2 \bigl(-1 + \bar\rho'\bigr) + 6 r f \bigl(f \
U' (1 + \bar\rho') + h_1 (5 + \bar\rho')\bigr) -  b \
\bigl(f (2 f (6 U (l (1 + l) \nonumber\\
&&- 16 \pi r^2 (P + \
\bar\rho)) + r (-8 + l + l^2 + 8 \pi r^2 (-7 P + \
\bar\rho)) U') - 3 r^2 k (-1 + \bar\rho') + 2 r (-6 r \
\omega^2 U + f (4 + l + l^2 \nonumber\\
&&+32 \pi r^2 P) U') \
\bar\rho') + r h_1 (-4 r^2 \omega^2 (-1 + \bar\rho') + \
f (16 + 11 l (1 + l) + (8 + l + l^2) \bar\rho' + 32 \pi r^2 \
(P - 3 \bar\rho + 2 P \bar\rho')))\bigr) \nonumber\\
&&+ \bigl(b\
\bigr)^2 \bigl(4 r^4 \omega^2 k (-1 + \bar\rho') + 2 \
(f)^2 (2 U (1 + l + l^2 - 8 \pi r^2 \bar\rho) (l (1 + l) - \
16 \pi r^2 (P + \bar\rho)) + r U' (-3 (1 + l + l^2)\nonumber\\
&&+ 8 \
\pi r^2 (- P (3 + 2 l (1 + l) + 8 \pi r^2 P) + (3 + l + l^2 + \
8 \pi r^2 P) \bar\rho - 8 \pi r^2 (\bar\rho)^2) + \
\bar\rho' \nonumber\\
&&+ (l + l^2 + 8 \pi r^2 P (2 + l + l^2 + 8 \pi r^2 \
P)) \bar\rho')) + r f (h_1 (1 + 8 \pi r^2 P) (2 + 3 l \
(1 + l) + 16 \pi r^2 (-2 \bar\rho + P (-1 + \bar\rho')) \nonumber\\
&&+ \
(2 + l + l^2) \bar\rho') + 2 r (- k (l + l^2 + 16 \pi r^2 \
P) (-1 + \bar\rho') + 2 \omega^2 U (8 \pi r^2 (P + \
\bar\rho) -  (1 + l + l^2 + 8 \pi r^2 P) \
\bar\rho')))\bigr)\bigr\}\nonumber\\
&&/\{4 r^2 f^2 \bigl(3 -  b \
(1 + l + l^2 - 8 \pi r^2 \bar\rho)\bigr)\},\\
&&k'=- \bigl\{32 \pi r \omega^2 b U \bigl(P + \bar\rho\bigr) -  \Bigr[k 
\bigl(-6 f + (b)^2 f (1 + 8 \pi r^2 P) (- l (1 + l) + 8 \pi r^2 P + 8 
\pi r^2 \bar\rho)\nonumber\\
&&+ b (4 r^2 \omega^2 + f (6 + l + l^2 - 8 \pi r^2 P 
- 24 \pi r^2 \bar\rho))\bigr)\Bigl]/[r f] -  \bigl[h1 \bigl(4 r^2 \omega^2 
+ f (l + l^2 - 16 \pi r^2 P - 16 \pi r^2 \bar\rho) (-2 \nonumber\\
&&+ b (l + l^2 - 
8 \pi r^2 P - 8 \pi r^2 \bar\rho))\bigr)\bigr]/[r^2 f] - 16 \pi f \bigl(P + 
\bar\rho\bigr) \bigl(2 + b (- l (1 + l) + 8 \pi r^2 P\nonumber\\
&&+ 8 \pi r^2 
\bar\rho)\bigr) U'\bigr\}/\bigl\{2 \bigl(3 -  b (1 + l + l^2 - 8 \pi r^2
\bar\rho)
\bigr)\bigr\}\\
&&h_1'=\bigl\{6 f h1 -  r \bigl(b\bigr)^3 f k \bigl(1 + 8 \pi r^2 P\bigr)^2 - 
 \bigl(b\bigr)^2 \bigl(4 r^3 \omega^2 k -  f (2 r (16 \pi r^2 
\omega^2 U (P + \bar\rho) + k (1 + 2 l + 2 l^2 + 16 \pi r^2 P - 8 \pi 
r^2 \bar\rho)) \nonumber\\
&&+ h1 (2 + 3 l + 3 l^2 - 128 \pi^2 r^4 (P)^2 - 8 (5 + l 
+ l^2) \pi r^2 \bar\rho + 64 \pi^2 r^4 (\bar\rho)^2 - 8 \pi r^2 P (1 
- 2 l - 2 l^2 + 24 \pi r^2 \bar\rho))) \nonumber\\
&&+ 16 \pi r (f)^2 (P + \bar\rho) (2 U (1
+ l + l^2 - 8 \pi r^2 \bar\rho) + r (1 + 8 \pi r^2 P) U')
\bigr) + b \bigl(- h1 (4 r^2 \omega^2 + f (8 + 3 l + 3 l^2 \nonumber\\
&&+ 8 \pi 
r^2 P - 56 \pi r^2 \bar\rho)) + r f (-9 k + 16 \pi f (P + \bar\rho) 
(6 U + r U'))\bigr)\bigr\}/\{2 r f \bigl(3 -  b (1 + l + l^2 - 8 \pi r^2 \bar
\rho)\bigr)\},
\eea
where we omitted the radial dependence of the functions.

These equations can be further written in terms of the Regge-Wheeler Master
function $X$ by using the following change of function
\bea
\label{khtoQ}
k&=&\bigl\{\bigl(l (1 + l) (-24 M^2 + 12 M r + (-1 + l) l (1 + l) (2 + l) \
r^2) - 24 M r^3 \omega^2\bigr) X \nonumber\\
&&+ 2 l \bigl(1 + l\bigr) r \bigl(-2 M \
+ r\bigr) \bigl(6 M + (-2 + l + l^2) r\bigr) X'\bigr\}/\bigl\{2 l \bigl(1 + \
l\bigr) \bigl(-2 + l + l^2\bigr) r^3\bigr\},\\
h_1&=&\{\bigl(-72 M^3 + 12 (3 + l + l^2) M^2 r -  (-1 + l) l (1 + l) (2 + 
l) r^3 + 3 M r^2 (l (1 + l) (-4 + l + l^2) + 4 r^2 \omega^2)\bigr) 
X \nonumber\\
&&-  \bigl(2 M -  r\bigr) r \bigl(6 M -  l (1 + l) r\bigr) 
\bigl(6 M + (-2 + l + l^2) r\bigr) X'\bigl\}/\bigr\{l \bigl(1 
+ l\bigr) \bigl(-2 + l + l^2\bigr) r^2 \bigl(-2 M + r\bigr)\}.
\eea

This transformation has the advantage to turn the vacuum equation ($U=0$) to
the simple Regge-Wheeler equation (in tortoise coordinates)
\be
X''(r_*)+\left( \frac{(r-2 M ) (6 M - l (1 + l) r)}{r^4} + \omega^2\right)
X(r_*)=0,
\ee
where $r_*$ is the tortoise coordinate defined by
\be
r_* = r +2M \log \left( \frac{r}{2M}-1 \right).
\ee

Note that the Regge-Wheeler equation originally describes the odd sector of the
metric perturbations. The system of equations describing the even sector is
called the Zerilli equation. However, these sectors are isospectral and are
actually equivalent, by virtue of the transformation \eqref{khtoQ}. Here we use the
Regge-Wheeler equation since the form of it is of Heun's equation
and is better suited for the construction of the analytic series solutions
of \cite{Mano:Suzuki:Takasugi:1996}.

\subsubsection{Boundary Conditions and Series}
Within the NS interior it is difficult to extend the numeric integration up to the boundary points
$r=0$ and $r=R$. This problem is solved by terminating the numeric integration very
close to these points and use analytic series solutions to extend the numeric solutions to
the boundaries. Furthermore, certain boundary conditions must be fulfilled, which are directly
implemented into the series solutions here.

At the origin $r=0$, the relevant boundary condition is simply the regularity of the
perturbation master functions. This imposes two independent conditions, so we
need only two (of four) integration constants to parametrize the solution around
$r=0$. The leading order series solutions read
\begin{align}
U(r) &= U_0 r^l [ 1 + \Order(r) ] , \\
k(r) &= k_0 r^l [ 1 + \Order(r) ] , \\
h_1(r) &= - \frac{2 r^{1 + l}}{1 + l} \bigl(k_0 - 8 \pi G U_0 f(0) [ P(0) + \bar\rho(P(0)) ] + \Order(r) \bigr) ,
\end{align}
where $U_0$ and $k_0$ are the integration constants.

The boundary condition at the surface is given by the requirement that the Lagrangian (comoving)
perturbation of the pressure or density vanishes for $r=R$, or explicitly

\begin{align}
\left.U'\right|_R &=-\frac{R b}{D} k\Bigl(3 f
                + b^2 f \hat P^2
                + b \bigl(4 R^2 \omega^2
                - 2 f \bigl((l+2) (l-1) + 2\hat P\bigr)\bigr)\Bigr)
                + \frac{2 R \omega^2 b U}{f( b\hat P-1)}
\nnl
        +\frac{ h_1}{D}\Bigl( -6 f
                -  b^2 f  \hat P
\bigl(2\hat P  + l \bigl(1 + l\bigr) \bigr) 
                + b \bigl(-4 R^2 \omega^2 + f \bigl(8\hat P + l (1 + l)
\bigr)\bigr)\left.\Bigr)\right|_R  ,\nonumber\\
D&=2 f^2 (b\hat P-1) \bigl(b ( \hat P + l (1 + l))-3\bigr)\ \ ,\ \hat
P=1+8 \pi G R^2 P. 
\end{align}

However, the behavior of the perturbations near the surface $r=R$ crucially depends on the EOS.
In the following analysis, we restrict to the case that the EOS near the surface is a polytrope
with index $1 \leq n < \infty$. Then the boundary condition is actually equivalent to regularity
of the perturbation master functions. The boundary condition allows us to eliminate one of the
four integration constants, so we are left with $U(R)$, $k(R)$, and $h_1(R)$. At the surface the
functions $k$ and $h_1$ must be continuous, which provides two further boundary conditions.
We are therefore able to express $k(R)$ and $h_1(R)$ in terms of the RW function $X(R)$ and its
derivative $X'(R)$ (which describe the exterior perturbation). The boundary
series finally reads
\begin{align}
U(r) &= U_R \nnl
        + (R-r) \Biggl( \frac{R^3 \omega^2 U_R}{M(2 M -  R)}
        - \frac{X_R}{4 l (1 + l) (-2 + l + l^2) M R
(-2 M + R)^2}
                (288 M^4 - 48 (3 + l + l^2) M^3 R \nnlq
                + (-1 + l) l (1 + l) (2 + l) R^4
(l + l^2 - 2 R^2 \omega^2) + 24 M^2 R^2 (l + l^2 + R^2
\omega^2) \nnlq
                - 2 M R^3 ((-1 + l) l (1 + l) (2 +
l) (1 + l + l^2) + 12 R^2 \omega^2)) \nnlq
        + \frac{X'_R}{2 l (1 + l) (-2 + l + l^2) M
(2 M -  R)}
                (-72 M^3 + 24 M^2 R + (-1 + l) l (1 +
l) (2 + l) R^3 \nnlq
                + M R^2 (- (-1 + l) l (1 + l) (2 +
l) - 12 R^2 \omega^2)) \Biggr) \nnl
        + (R-r)^2 \Biggl( \frac{(- l (1 + l) M (2 M - 
R) + (M - 2 R) R^3 \omega^2) U_R}{M R (-2 M +
R)^2} \nnlq
                + \frac{X_R}{2 l (1 + l) (-2 + l + l^2)
M (2 M -  R)^3 R^2}
                (576 M^5 - 96 (3 + l + l^2) M^4 R \nnlq
                - 3 (-1 + l) l (1 + l) (2 + l)
(2 + l + l^2) M R^4
                + 2 (-12 + (-1 + l) l (1 + l) (2 +
l)) M R^6 \omega^2 \nnlq
                + (-1 + l) l (1 + l) (2 + l) R^5
(l + l^2 - 2 R^2 \omega^2)
                - 12 M^3 R^2 (l (1 + l) (-10 + 3 l (1 +
l)) + 14 R^2 \omega^2) \nnlq
                + 2 M^2 R^3 ((-1 + l) l (1 + l) (2
+ l) (13 + l + l^2) + 48 R^2 \omega^2)) \nnlq
        + \frac{X'_R}{l (1 + l) (-2 + l + l^2) M R
(-2 M + R)^2}
                (144 M^4 - 48 M^3 R -  (-1 + l) l (1 +
l) (2 + l) R^4 \nnlq
                + M^2 R^2 (-7 (-1 + l) l (1 + l)
(2 + l) - 12 R^2 \omega^2) + 4 M R^3 ((-1 + l) l
(1 + l) (2 + l) + 3 R^2 \omega^2)) \Biggr) \nnl
+ \Order\left[(R-r)^3\right] , \\
k(r) &= \frac{(l (1 + l) (-24 M^2 + 12 M R + (-1 +
l) l (1 + l) (2 + l) R^2) - 24 M R^3
\omega^2) X_R}{2 l (1 + l) (-2 + l + l^2) R^3}
        -  \frac{(2 M -  R) (6 M + (-2 + l + l^2)
R) X'_R}{(-2 + l + l^2) R^2} \nnl
+ (R-r) \Biggl(\Bigl(- \frac{l (1 + l)}{2 R^2}
        + \frac{(6 M + (-2 + l + l^2) R) \omega^2}{(-2 + l +
l^2) (-2 M + R)}\Bigr) X_R 
        + \Bigl(- \frac{l (1 + l)}{2 R} + \frac{12 M \omega^2}{l
(1 + l) (-2 + l + l^2)}\Bigr) X'_R\Biggr) \nnl
+ \Order\left[(R-r)^2\right] , \\
h_1(r) &= \frac{X_R}{l \left(1 + l\right) \left(-2 + l + l^2\right) R^2 \left(-2
M + R\right)}
        [ -72 M^3 + 12 (3 + l + l^2) M^2 R \nnlq
        -  (-1 + l) l (1 + l) (2 + l) R^3
        + 3 M R^2 (l (1 + l) (-4 + l + l^2) + 4
R^2 \omega^2) ] \nnlq
        + \frac{\left(6 M -  l (1 + l) R\right) \left(6 M +
(-2 + l + l^2) R\right) X'_R}{l \left(1 + l\right) \left(-2 + l +
l^2\right) R} \nnl
        + (R-r) \Bigl(\frac{X_R}{l (1 + l) (-2 + l +
l^2) R^3 (-2 M + R)^2}
                (-144 M^4 + 24 (3 + 2 l (1 + l)) M^3
R \nnlq
                + (-1 + l) l (1 + l) (2 + l)
R^4 (l + l^2 -  R^2 \omega^2)
                + 12 M^2 R^2 (l (1 + l) (-5 + l + l^2)
+ 7 R^2 \omega^2) \nnlq
                -  M R^3 (l (1 + l) (-16 + l (1 +
l) (1 + 2 l (1 + l))) + 24 R^2 \omega^2)) \nnlq
        + \frac{X'_R}{l (1 + l) (-2 + l + l^2) (2
M -  R) R^2}
                (-72 M^3 + 12 (2 + l + l^2) M^2 R \nnlq
                - 2 (-1 + l) l (1 + l) (2 +
l) R^3 + M R^2 (l (1 + l) (-20 + 7 l (1 +
l)) + 12 R^2 \omega^2)) \Bigr) \nnl
+ \Order\left[(R-r)^2\right] ,
\end{align}
where $U_R,X_R,X'_R$ are the functions $U,X,X'$ evaluated at the radius $R$.

If the RW equation is solved numerically using a direct integration method
\cite{Chandrasekhar:Detweiler:1975}, then one must derive series solutions
for the RW function $X$ at $r=\infty$ in a similar manner (and for BH also
at the horizon, where the physical boundary condition only permits an
ingoing flux).

\subsection{Analytic Solutions \texorpdfstring{from \cite{Mano:Suzuki:Takasugi:1996}}{}\label{MST}}

\subsubsection{Solutions to the RW Equation}
The pair of independent (UV or "near-zone") solutions $\{ X_0^{\nu}, X_0^{-\nu-1} \}$ and
the pair of independent (IR or "far-zone") solutions $\{ X_C^{\nu}, X_C^{-\nu-1} \}$ are
given by \cite[Eqs.\ (2.16) and (3.6)]{Mano:Suzuki:Takasugi:1996}, see also
\cite{Leaver:1986:1, Sasaki:Tagoshi:2003},
\begin{align}
X_0^{\nu} &= e^{i (x-1) \epsilon } (-x)^{-i \epsilon } (1-x)^{\nu +i \epsilon +1}
\sum_{n=-\infty}^{\infty} (1-x)^n a_n^{\nu} \frac{\Gamma (2 n+2 \nu +1) \Gamma (-n-i \epsilon -\nu -2)}{\Gamma (n-i \epsilon +\nu +3)} \nnlq
        \times {}_2F_1(-n-i \epsilon -\nu -2,-n-i \epsilon -\nu +2;-2 n-2 \nu ; 1 / (1-x) ) \\
X_C^{\nu} &= \left(1-\frac{\epsilon }{z}\right)^{-i \epsilon }
\sum_{n=-\infty}^{\infty} 2^{\nu +n} i^n e^{-i z} z^{\nu +n+1} a_n^{\nu} \frac{\Gamma (n-i \epsilon +\nu -1) \Gamma (n-i \epsilon +\nu +1)}{\Gamma (2 (n+\nu )+2) \Gamma (n+i \epsilon +\nu +3)} \nnlq
        \times {}_1F_1(n+i \epsilon+\nu +1;2 (n+\nu )+2;2 i z)
\end{align}
where
\begin{align}
\epsilon &= 2 M \omega , \\
z &= \omega r , \\
x &= 1 - \frac{r}{2 M} = 1 - \frac{z}{\epsilon} , \\
{}_2F_1(a, b; c; z) &= \sum_{n=0}^{\infty} \frac{(a)_n (b)_n}{(c)_n} \frac{z^n}{n!} , \\
{}_1F_1(a; c; z) &= \sum_{n=0}^{\infty} \frac{(a)_n}{(c)_n} \frac{z^n}{n!} , \\
(a)_n &= \frac{\Gamma(a+n)}{\Gamma(a)} .
\end{align}
Here $\Gamma$ is the Gamma Function, ${}_2F_1$ is the Gauss Hypergeometric Function, ${}_1F_1$ is the
Confluent Hypergeometric Function, and $(a)_n$ is the Pochhammer function (rising factorial). The
coefficients $a_n^{\nu}$ entering the series and the renormalized angular momentum $\nu$ are
explained in the next section.

The relation between the solutions is given by \cite[Eq.\ (4.1)]{Mano:Suzuki:Takasugi:1996}
\begin{equation}
X_0^{\nu} = K_{\nu} X_C^{\nu} ,
\end{equation}
where \cite[Eq.\ (4.2)]{Mano:Suzuki:Takasugi:1996}
\begin{align}
K_{\nu} &= -\frac{\pi  i^r 2^{-\nu -r} \epsilon ^{-\nu -r-1} \csc (\pi  (\nu +i \epsilon ))}{\Gamma (r+i \epsilon +\nu -1) \Gamma (r+i \epsilon +\nu +1) \Gamma (r+i \epsilon +\nu +3)}
   \left[ \sum_{n=r}^{\infty} a_n^{\nu} \frac{\Gamma (n+r+2 \nu +1) \Gamma (n+i \epsilon +\nu -1)}{(n-r)! \Gamma (n-i \epsilon +\nu +3)} \right] \nnl
   \times \left[ \sum_{n=-\infty}^r a_n^{\nu} \frac{\Gamma (n-i \epsilon +\nu -1) \Gamma (n-i \epsilon +\nu +1)}{(r-n)! \Gamma (n+r+2 \nu +2) \Gamma (n+i \epsilon +\nu +1)
   \Gamma (n+i \epsilon +\nu +3)} \right]^{-1} . \label{KMST}
\end{align}
The value of $r \in \mathbb{Z}$ is in principle arbitrary, which can also be checked numerically.
For definiteness, we chose $r=0$.

\subsubsection{Recurrence Relation for \texorpdfstring{$a_n^{\nu}$}{a(n)}}
The three-term recurrence relation for the $a_n^{\nu}$ reads \cite[Eq.\ (2.5)]{Mano:Suzuki:Takasugi:1996}
\begin{equation}
\alpha_n^{\nu} a_{n+1}^{\nu} + \beta_n^{\nu} a_n^{\nu} + \gamma_n^{\nu} a_{n-1}^{\nu} = 0 ,
\end{equation}
where the coefficients are given by \cite[Eq.\ (2.6), (2.7), and (2.8)]{Mano:Suzuki:Takasugi:1996}
\begin{align}
\alpha_n^{\nu} &= -\frac{i \epsilon  (\nu +n-i \epsilon -1) (\nu +n-i \epsilon +1) (\nu +n+i \epsilon -1)}{(\nu +n+1) (2 (\nu +n)+3)} , \\
\beta_n^{\nu} &= -l (l+1)+(\nu +n) (\nu +n+1)+\frac{\left(\epsilon ^2+4\right) \epsilon ^2}{(\nu +n) (\nu +n+1)}+2 \epsilon ^2 , \\
\gamma_n^{\nu} &= \frac{i \epsilon  (\nu +n-i \epsilon +2) (\nu +n+i \epsilon ) (\nu +n+i \epsilon +2)}{(\nu +n) (2 (\nu +n)-1)} .
\end{align}
We proceed along the lines of \cite{Mano:Suzuki:Takasugi:1996, Sasaki:Tagoshi:2003} by
defining continued fractions $R_n(\nu)$ and $L_n(\nu)$ \cite[Eq.\ (2.9) and (2.10)]{Mano:Suzuki:Takasugi:1996}
\begin{align}
R_n(\nu) &= \frac{a_n^{\nu}}{a_{n-1}^{\nu}}
        = - \frac{\gamma_n^{\nu}}{\beta_n^{\nu} + \alpha_n^{\nu} R_{n+1}(\nu)} , \label{Rcf} \\
L_n(\nu) &= \frac{a_n^{\nu}}{a_{n+1}^{\nu}}
        = - \frac{\alpha_n^{\nu}}{\beta_n^{\nu} + \gamma_n^{\nu} L_{n-1}(\nu)} .
\end{align}
From these expressions it is straightforward to infer that
\begin{align}
\lim_{n \rightarrow \infty} n R_n(\nu) &= -\frac{i \epsilon}{2} , \label{Rlim} \\
\lim_{n \rightarrow - \infty} n L_n(\nu) &= \frac{i \epsilon}{2} , \label{Llim}
\end{align}
provided that the continued fractions converge in the specified limit. The corresponding
solution to the three-term recurrence relation is called the minimal solution in the specific limit and is
guaranteed to exist. But the minimal solutions for $n \rightarrow \infty$ and
$n \rightarrow -\infty$ are not necessarily the same, e.g., in general one can fulfill
either (\ref{Rlim}) or (\ref{Llim}), but not both at the same time. However,
requiring both (\ref{Rlim}) and (\ref{Llim}) fixes the renormalized angular momentum $\nu$.
This is dictated by the convergence of the analytic solutions to the RW equation.

In practice, one uses the limit (\ref{Rlim}) as a starting value for $R_n(\nu)$ at some large but finite $n>0$.
From the continued fraction (\ref{Rcf}) one can then easily determine $R_n(\nu)$ for any smaller $n$.
An analogous process can be applied to $L_n(\nu)$, this time starting from a large but finite negative $n<0$.
Finally, one imposes the consistency condition \cite[Eq.\ (2.11)]{Mano:Suzuki:Takasugi:1996}
\begin{equation}
R_n(\nu) L_{n-1}(\nu) = 1 ,
\end{equation}
at some value for $n$. For definiteness, we chose $n=1$. We solve this condition for $\nu$ using
standard numerical root-finding procedures starting from the initial value \cite[Eq.\ (6.3)]{Mano:Suzuki:Takasugi:1996}
\begin{equation}
\nu = l
  + \left(-\frac{(l-2)^2 (l+2)^2}{2 l (2 l-1) (2 l+1)}-\frac{4}{l (l+1)}+\frac{(l-1)^2 (l+3)^2}{(2 l+1) (2 l+2) (2 l+3)}-2\right) \frac{\epsilon ^2}{2 l+1}
  + \Order(\epsilon^4) .
\end{equation}
Finally, one can determine $a_n^{\nu}$, which is fixed up to an overall factor. As in
\cite{Mano:Suzuki:Takasugi:1996}, we set $a_0^{\nu} = 1$.
Besides numeric approaches, it is of course possible to work out analytic series expansions
in $\epsilon$ for $\nu$ and $a_n^{\nu}$, see \cite[Sec.\ 6]{Mano:Suzuki:Takasugi:1996}
for further discussions.

\subsubsection{Normalization of Analytic Solutions and Wronskian\label{normMST}}
It is straightforward to determine the asymptotic behavior of $X_C^{\nu}$ as
\begin{equation}
X_C^{\nu} \stackrel{r_* \rightarrow \infty}{\sim}
        A_{C \, \text{in}}^{\nu} e^{-i \omega r_*} + A_{C \, \text{out}}^{\nu} e^{i \omega r_*} ,
\end{equation}
with the complex amplitudes
\begin{align}
A_{C \, \text{in}}^{\nu} &= \frac{1}{2} i^{-\nu +i \epsilon -1} \sum_{n = - \infty}^{\infty}
        - i^n (2\epsilon)^{-i \epsilon } e^{i \pi  \left(\nu +\frac{n}{2}\right)} a_n^{\nu} \frac{\Gamma (n-i \epsilon
   +\nu -1)}{\Gamma (n+i \epsilon +\nu +3)} , \label{N1MST} \\
A_{C \, \text{out}}^{\nu} &= \frac{1}{2} i^{-\nu +i \epsilon -1} \sum_{n = - \infty}^{\infty}
        (2\epsilon)^{i \epsilon } a_n^{\nu} \frac{\Gamma (n-i \epsilon +\nu -1) \Gamma (n-i \epsilon +\nu +1)}{\Gamma
   (n+i \epsilon +\nu +1) \Gamma (n+i \epsilon +\nu +3)} . \label{N2MST}
\end{align}
Then we obtain for the normalization 
\begin{equation}\label{NMST}
N_{\nu} = \frac{1}{2} (A_{C \, \text{in}}^{\nu} A_{C \, \text{out}}^{\nu})^{-\frac{1}{2}} .
\end{equation}
Notice that $N_{\nu}$ is multivalued (bi-valued). Eventually the second root must be
used in some frequency regimes in order to make results continuous.
In the present investigation this is necessary at about $\omega R / 2 \pi > 0.23$.
It follows that
\begin{align}
X_N^{\nu} &:= N_{\nu} X_C^{\nu} , \\
&\stackrel{r_* \rightarrow \infty}{\sim}
        \frac{1}{2} \left[ \left( \frac{A_{C \, \text{out}}^{\nu}}{A_{C \, \text{in}}^{\nu}} \right)^{-\frac{1}{2}} e^{-i \omega r_*}
        + \left( \frac{A_{C \, \text{out}}^{\nu}}{A_{C \, \text{in}}^{\nu}} \right)^{\frac{1}{2}} e^{i \omega r_*} \right] , \\
&=\frac{1}{2} \left[ e^{-i (\omega r_* + \alpha_{\nu})} + e^{i (\omega r_* + \alpha_{\nu})} \right] ,
\qquad \text{with } \alpha_{\nu} := \frac{1}{2 i} \log \frac{A_{C \, \text{out}}^{\nu}}{A_{C \, \text{in}}^{\nu}} , \\
&= \cos(\omega r_* + \alpha_{\nu}) ,
\end{align}
as envisaged. Because the RW equation has real coefficients, it is guaranteed that a real
solution for $\alpha_{\nu}$ exists.

It is also straightforward to obtain the Wronskian (\ref{Wrons}) from this analysis as
\begin{align}\label{WronsMST}
W_* &= 2 i \omega N_{\nu}N_{-\nu-1} ( A_{C \, \text{in}}^{\nu} A_{C \, \text{out}}^{-\nu-1}
        - A_{C \, \text{out}}^{\nu} A_{C \, \text{in}}^{-\nu-1} ),\\
&= \omega \sin(\alpha_{\nu} - \alpha_{-\nu-1})
\end{align}
The Wronskian (\ref{Wrons}) based on the solutions $X_{\delta_l}^{\nu}$
and $X_{\delta_l}^{-\nu-1}$ follows by expanding (\ref{WronsMST}) in
$\epsilon$ for generic $l$.

\subsection{Effective Source\label{effS}}

\subsubsection{Inhomogeneous RW Equation}
The homogeneous Zerilli and Regge-Wheeler equations describe vacuum perturbations.
Considering additional matter fields will source the vacuum perturbation
equations. If the right hand side of the perturbed Einstein equation is
$\delta T_{\mu\nu}$, the combination of the metric perturbations leading to the
source to Zerilli equation and further transformed to Regge-Wheeler equation is 
 \bea
 -S&=&\frac{2 \ell_2 r^2 \bigl(-2 M + r\bigr)^2 \bigl(6 (-2 + \ell_2) M + (4 - 
2 \ell_2 + (-2 + \ell_2)^{\tfrac{1}{2}} \ell_2^{\tfrac{1}{2}} ((-2 + \ell_2) 
\ell_2)^{\tfrac{1}{2}}) r\bigr) }{-864 M^4 - 144 \bigl(-5 + 
\ell_2\bigr) M^3 r + 36 \bigl(-4 + \ell_2^2\bigr) M^2 r^2 + 12 \bigl(-2 + 
\ell_2\bigr)^2 \ell_2 M r^3 + \bigl(-2 + \ell_2\bigr)^3 \ell_2 r^4}\mathcal
T_{00}\nonumber\\
&&-  \frac{2 
\sqrt{2} \ell_2 \bigl(-4 + 2 \ell_2 -  (-2 + \ell_2)^{\tfrac{1}{2}} 
\ell_2^{\tfrac{1}{2}} ((-2 + \ell_2) \ell_2)^{\tfrac{1}{2}}\bigr) r^2 
\bigl(-2 M + r\bigr)^2 }{\bigl(6 M + (-2 + \ell_2) r\bigr) 
\bigl(-144 M^3 + 72 M^2 r + 6 (-2 + \ell_2) \ell_2 M r^2 + (-2 + \ell_2)^2
\ell_2 
r^3\bigr) \omega}\mathcal T_{01}\nonumber\\
&&-  \frac{4 \sqrt{2} \bigl(-2 + 
\ell_2\bigr)^{\tfrac{1}{2}} \bigl((-2 + \ell_2) \ell_2\bigr)^{\tfrac{1}{2}} 
r \bigl(-2 M + r\bigr)^2 \bigl(-12 M^2 + 12 M r + (-2 + \ell_2) 
r^2\bigr) }{\bigl(6 M + (-2 + \ell_2) r\bigr) \bigl(-144 M^3 + 72 
M^2 r + 6 (-2 + \ell_2) \ell_2 M r^2 + (-2 + \ell_2)^2 \ell_2 r^3\bigr) \omega}
\mathcal T_{0e}\nonumber\\
&&-  
\frac{2 \ell_2 r^4 \bigl(6 (-2 + \ell_2) M + (4 - 2 \ell_2 + (-2 + 
\ell_2)^{\tfrac{1}{2}} \ell_2^{\tfrac{1}{2}} ((-2 + \ell_2) 
\ell_2)^{\tfrac{1}{2}}) r\bigr) }{-864 M^4 - 144 \bigl(-5 + 
\ell_2\bigr) M^3 r + 36 \bigl(-4 + \ell_2^2\bigr) M^2 r^2 + 12 \bigl(-2 + 
\ell_2\bigr)^2 \ell_2 M r^3 + \bigl(-2 + \ell_2\bigr)^3 \ell_2 r^4}\mathcal
T_{11} \nonumber\\
&&+ \frac{2 
\sqrt{2} \bigl(-2 + \ell_2\bigr)^{\tfrac{1}{2}} \bigl((-2 + \ell_2) 
\ell_2\bigr)^{\tfrac{1}{2}} \bigl(2 M -  r\bigr) r^4 }{-144 M^3 + 
72 M^2 r + 6 \bigl(-2 + \ell_2\bigr) \ell_2 M r^2 + \bigl(-2 + \ell_2\bigr)^2 
\ell_2 r^3}\mathcal T_{1e}\nonumber\\
&&-  \frac{2 \sqrt{2} 
\bigl(-2 + \ell_2\bigr)^{\tfrac{1}{2}} \bigl((-2 + \ell_2) 
\ell_2\bigr)^{\tfrac{1}{2}} \bigl(2 M -  r\bigr)^3 r^2}{\bigl(144 M^3 - 72 M^2 r
- 6
(-2 + \ell_2) \ell_2 M r^2 -  (-2 + 
\ell_2)^2 \ell_2 r^3\bigr) \omega} \mathcal T_{0e}'\nonumber\\
&&-  \frac{2 \sqrt{2} \bigl((-2 + \ell_2) 
\ell_2\bigr)^{\tfrac{1}{2}} \bigl(2 M -  r\bigr) r^4 \bigl(6 M + (-2 
+ \ell_2) r\bigr) }{-144 M^3 + 72 M^2 r + 6 \bigl(-2 + \ell_2\bigr) \ell_2 
M r^2 + \bigl(-2 + \ell_2\bigr)^2 \ell_2 r^3} \mathcal T_e,
 \eea
where we introduced the notation $\ell_2=l(l+1)$ and where $\mathcal T_\mathcal{Z}$,
$\mathcal{Z}\in\{00,01,11,{0e},{0o},{1e},{1o},{t},{e},{o}\}$ are the Zerilli tensor
spherical harmonic (TSH) components of $T^{\mu\nu}$ defined by
\be
\mathcal T_\mathcal{Z}  = N_\mathcal{Z} \int T^{\mu\nu} Y_{\mathcal{Z},\mu\nu}^* d\Omega .
\ee
Here $Y_{\mathcal{Z},\mu\nu}$ are the Zerilli TSH \cite{Zerilli:1970:2} and
$N_\mathcal{Z}$ their normalizations given by $N=\{1,-1,1,-1,-1,1,1,1,1,1\}$.

Finally, the sourced Zerilli equation converted to Regge-Wheeler form is given
by
\be
X''(r_*)+\left( \frac{(r-2 M ) (6 M - \ell_2 r)}{r^4} + \omega^2\right)
X(r_*)=S.
\ee

\subsubsection{Stress Tensor}
The stress tensor up to the quadrupole approximation reads
\cite{Steinhoff:Puetzfeld:2009, Steinhoff:2011}
\begin{equation}
\sqrt{-g} T^{\mu\nu} = \int d \tau \bigg[u^{(\mu} p^{\nu)} \delta_{(4)} + \frac{1}{3} R_{\alpha\beta\gamma}{}^{(\mu} J^{\nu)\gamma\beta\alpha} \delta_{(4)}
  - \nabla_\alpha ( S^{\alpha(\mu} u^{\nu)} \delta_{(4)} )- \frac{2}{3} \nabla_\beta \nabla_\alpha ( J^{\beta(\mu\nu)\alpha} \delta_{(4)} ) \bigg] ,
\end{equation}
where
\begin{equation}
  p^\mu = m u^\mu - \frac{\delta S^{\mu\nu}}{d s} u_\nu + \frac{4}{3} u_{b} R_{cde}{}^{[\mu} J^{b]edc},
\end{equation}
and $\delta_{(4)} = \delta(x^\mu-z^\mu)$.
Here the 4-quadrupole $J^{\alpha\beta\mu\nu}$ has the same symmetries as the Riemann tensor
\begin{eqnarray}
  J^{\alpha\beta\mu\nu} = J^{[\alpha\beta][\mu\nu]} = J^{\mu\nu\alpha\beta}, \\
  J^{[\alpha\beta\mu]\nu} = 0 \quad \Leftrightarrow \quad	J^{\alpha\beta\mu\nu} + J^{\beta\mu\alpha\mu} + J^{\mu\alpha\beta\nu} = 0.
\end{eqnarray}
It results directly from the effective Lagrangian as \cite{Bailey:Israel:1975}
\begin{equation}\label{Jdef}
  J^{\alpha\beta\mu\nu} = - 6 \frac{\partial L_{\text{int}}}{\partial R_{\alpha\beta\mu\nu}},
\end{equation}
which is defined by (\ref{Seff}),
\begin{equation}
S_{\text{eff}} = \int d \tau \, L_{\text{int}} , \qquad
 L_{\text{int}} = \left[ - m - \frac{1}{2} E_{\mu\nu} e_a{}^{\mu} e_b{}^{\nu} Q^{ab} + \dots \right] .
\end{equation}
Here $e_a{}^{\mu}$ is the tetrad defining the local frame. We formally
extend the local spatial indices $a$, $b$ by a time component here. This is fine if we also set
all time components of quantities defined in the local frame to zero, e.g., $Q^{a(0)} = 0$.
For the sake of the variation, we can then consider $e_a{}^{\mu}$ as unconstrained.
(We implement the constraint $e^{(0)\mu} =  u^{\mu}$ at the level of the equations of motion.)
Notice that the result (\ref{Jdef}) from \cite{Bailey:Israel:1975} is valid in the presence of
a tetrad $e_a{}^{\mu}$. Using $E_{\mu\nu} = R_{\mu\alpha\nu\beta} u^\mu u^\nu$ (in vacuum) we get
\begin{equation}
  J^{\alpha\beta\mu\nu} = - 3 u^{[\alpha} Q^{\beta][\mu} u^{\nu]} ,
\end{equation}
which is what we anticipated. The spin vanishes here, $S^{\mu\nu}=0$.
We further disregard the mass term $m$, as we are only interested in the
contributions from the quadrupole here.

\subsubsection{Quadrupole Source for RW Equation}
For technical reasons we are not working with a local Cartesian basis, but in one that is
adapted to TSH, i.e.,
\begin{align}
\eta^{ab}_{\text{TSH}} &= g_{\mu\nu} e^{a\mu} e^{b\nu} , \\
\eta^{ab}_{\text{TSH}} &= \text{diag}(-1, 1, 1, 1 / \sin^2 \theta) , \\
\eta_{ab}^{\text{TSH}} &= \text{diag}(-1, 1, 1, \sin^2 \theta) .
\end{align}
We can then transform components in this local basis to TSH components in the usual way.
Our choice for the frame field reads
\begin{equation}
\left( e^{a\mu} \right) = \left(
\begin{array}{cccc}
 -\frac{1}{\sqrt{f(r)}} & 0 & 0 & 0 \\
 0 & \frac{1}{\sqrt{b(r)}} & 0 & 0 \\
 0 & 0 & \frac{1}{r} & 0 \\
 0 & 0 & 0 & \frac{1}{r \sin^2\theta}
\end{array}
\right)
+ e^{-i \omega t} Y_{lm}(\Omega) \left(
\begin{array}{cccc}
\frac{h_0(r)}{2 \sqrt{f(r)}} & 0 & 0 & 0 \\
\frac{h_1(r)}{\sqrt{b(r)} f(r)} & \frac{h_2(r)}{2 \sqrt{b(r)}} & 0 & 0 \\
0 & 0 & \frac{k(r)}{2 r} & 0 \\
0 & 0 & 0 & \frac{k(r)}{2 r \sin^2\theta}
\end{array}
\right) .
\end{equation}
Obviously it fulfills $e^{(0)\mu} = u^{\mu}$.

The required components of the stress tensor in TSH basis $\mathcal{T}_X$ are given by
\begin{align}
\mathcal{T}_{00} &= \frac{\sqrt{3} \mathcal{Q}_{\text{1e}} \left(r (r-2 M) \delta _r'(r)+(2 r-3 M) \delta _r(r)\right)}{r^2 (2 M-r)}+\frac{\sqrt{3} \mathcal{Q}_{\text{e}} \delta _r(r)}{\sqrt{r^3 (r-2 M)}} \nnl
-\frac{\mathcal{Q}_{\text{t}} \left(r \left((5 r-7 M) \delta _r'(r)+r (r-2 M) \delta _r''(r)\right)+3 (M+2 r) \delta _r(r)\right)}{r^3 \sqrt{2-\frac{4 M}{r}}} \\
\mathcal{T}_{01} &= \frac{\mathcal{Q}_{\text{t}} \omega  \sqrt{r-2 M} \left(r \delta _r'(r)+3 \delta _r(r)\right)}{r^{3/2}}+\frac{\sqrt{\frac{3}{2}} \mathcal{Q}_{\text{1e}} \omega  \delta _r(r)}{r} \\
\mathcal{T}_{11} &= \frac{\mathcal{Q}_{\text{t}} \sqrt{r-2 M} \left(\delta _r(r) \left(-11 M^2+6 M r+r^4 \omega ^2\right)+M r (r-2 M) \delta _r'(r)\right)}{\sqrt{2} r^{9/2}}-\frac{\sqrt{3} M \mathcal{Q}_{\text{1e}} (2 M-r) \delta _r(r)}{r^4} \\
\mathcal{T}_{0e} &= \frac{\mathcal{Q}_{\text{e}} \omega  \delta _r(r)}{\sqrt{r^3 (r-2 M)}}-\frac{\sqrt{\frac{3}{2}} \mathcal{Q}_{\text{t}} \omega  \delta _r(r)}{\sqrt{r^3 (r-2 M)}}-\frac{\mathcal{Q}_{\text{1e}} \omega  \left(r \delta _r'(r)+3 \delta _r(r)\right)}{2 r^2} \\
\mathcal{T}_{1e} &= -\frac{\mathcal{Q}_{\text{1e}} \delta _r(r) \left(-6 M^2+3 M r+r^4 \omega ^2\right)}{2 r^5}+M \mathcal{Q}_{\text{e}} \sqrt{\frac{r-2 M}{r^9}} \delta _r(r)-\sqrt{\frac{3}{2}} M \mathcal{Q}_{\text{t}} \sqrt{\frac{r-2 M}{r^9}} \delta _r(r) \\
\mathcal{T}_{e} &= \frac{\mathcal{Q}_{\text{e}} \left(\delta _r(r) \left(M^2-r^4 \omega ^2\right)+M r (r-2 M) \delta _r'(r)\right)}{2 \sqrt{r^{11} (r-2 M)}}
\end{align}
where $\mathcal{Q}_X$ denotes the quadrupole in local frame TSH components.
 Here $\delta_r(r)$ is given by the right hand
side of (\ref{Riesz}), i.e.,
\begin{equation}
\delta_r(r) = ( r c_l )^{-\delta_l} \frac{\Gamma(\frac{3-\delta}{2})}{\pi^{3/2} 2^{\delta} \Gamma(\frac{\delta}{2})}
	\mu_0^{\delta} r^{\delta-3} .
\end{equation}

\subsubsection{Quadrupole Components\label{quadcomp}}
Finally, we must obtain the components of the quadrupole from (\ref{Qren}).
We actually work with the TSH version of (\ref{Qren}), but this does not pose
any problem. For $l=2$ the components needed for the present computation read
\begin{align}
\mathcal{Q}_{1e} &= - F(\omega) 2\pi\sqrt{3} \int_0^{\infty} \sqrt{1-\frac{2 M}{r}} \Bigg[
        \frac{\frac{M}{2 r}}{1-\frac{2 M}{r}}
                \left[r^2 \omega ^2 \left(2+\frac{3 M}{r}\right)+\frac{12 M}{r}-6\right] X
        - \left(2+\frac{3 M}{r}\right) r X'
        \Bigg] \frac{\delta_r(r)}{r} \, d r , \\
\mathcal{Q}_{t} &= - F(\omega) \frac{3\pi}{\sqrt{2}} \left(4+M^2 \omega ^2\right)
        \int_0^{\infty} X \frac{\delta_r(r)}{r} \, d r , \\
\mathcal{Q}_{e} &= - F(\omega) \sqrt{3}\pi \int_0^{\infty} \left[
        \frac{1-\frac{M}{r}}{1-\frac{2 M}{r}}
                \left[r^2 \omega ^2 \left(2+\frac{3 M}{r}\right)+\frac{12 M}{r}-6\right] X
        + \left[\frac{M}{r} \left(r^2 \omega^2+6\right)-2 \right] r X'
   \right] \frac{\delta_r(r)}{r} \, d r .
\end{align}
The angular integration was already performed. Remember that the Riesz kernel is
independent of angular coordinates.

\end{widetext}
}

\end{document}